\documentclass[draft]{agujournal2019}
\usepackage{url}
\usepackage{lineno}
\usepackage[inline]{trackchanges} 
\usepackage{soul}
\usepackage{amsmath}
\usepackage{xcolor}

\journalname{Journal of Geophysical Research: Planets}

\begin{document} 

   \title{Early Habitability and Crustal Decarbonation of a Stagnant-Lid Venus}

   \authors{D. H\"oning
          \affil{1,2},
         P. Baumeister
         \affil{3,4},
         J. L. Grenfell
         \affil{3},
         N. Tosi
         \affil{3},
         M. J. Way
         \affil{5,6,7},
          }
   \affiliation{1}{Origins Center, Nijenborgh 7, 9747 AG Groningen, The Netherlands}
   \affiliation{2}{
             Department of Earth- and Life Sciences, Vrije Universiteit Amsterdam, Amsterdam, The Netherlands}
   \affiliation{3}{
          Institute of Planetary Research, German Aerospace Center (DLR), Berlin, Germany}
   \affiliation{4}{
          Department of Astronomy and Astrophysics, Berlin Institute of Technology, Berlin, Germany
             }
   \affiliation{5}{
          NASA Goddard Institute for Space Studies, New York, USA
             }
   \affiliation{6}{
          GSFC Sellers Exoplanet Environments Collaboration
             }
   \affiliation{7}{
          Theoretical Astrophysics, Department of Physics and Astronomy, Uppsala University, Uppsala, SE-75120, Sweden
             }
  \correspondingauthor{Dennis H\"oning}{d.hoening@vu.nl}
 
 \begin{keypoints}
\item Stagnant-lid model scenarios of early Venus tuned to reproduce present-day observations suggest an early habitable period of up to 900 Myr
\item After runaway greenhouse, ongoing volcanism and crustal burial on stagnant-lid planets cause rapid crust decarbonation and raise of atmospheric CO$_2$
\item  Stagnant-lid planets have bimodal distribution of atmospheric CO$_2$: abundance is low on habitable worlds and high after runaway greenhouse
\end{keypoints}
 
\justifying
 
\begin{abstract}
Little is known about the early evolution of Venus and a potential habitable period during the first one billion years. In particular, it remains unclear whether or not plate tectonics and an active carbonate-silicate cycle were present. In the presence of liquid water but without plate tectonics, weathering would have been limited to freshly produced basaltic crust, with an early carbon cycle restricted to the crust and atmosphere. With the evaporation of surface water, weathering would cease. With ongoing volcanism, carbonate sediments would be buried and sink downwards. Thereby, carbonates would heat up until they become unstable and the crust would become depleted in carbonates. With CO$_2$ supply to the atmosphere the surface temperature rises further, the depth below which decarbonation occurs decreases, causing the release of even more CO$_2$.

We assess the habitable period of an early stagnant-lid Venus by employing a coupled interior-atmosphere evolution model accounting for CO$_2$ degassing, weathering, carbonate burial, and crustal decarbonation. We find that if initial surface conditions allow for liquid water, weathering can keep the planet habitable for up to 900 Myr, followed by evaporation of water and rapid crustal carbonate depletion. For the atmospheric CO$_2$ of stagnant-lid exoplanets, we predict a bimodal distribution, depending on whether or not these planets experienced a runaway greenhouse in their history. Planets with high atmospheric CO$_2$ could be associated with crustal carbonate depletion as a consequence of a runaway greenhouse, whereas planets with low atmospheric CO$_2$ would indicate active silicate weathering and thereby a habitable climate.
\end{abstract}

\section*{Plain Language Summary}

Today, Venus has a thick atmosphere mainly composed of CO$_2$ and a surface that is too hot for any liquid water to exist. However, four billion years ago, the Sun was much fainter, and if Venus’ atmosphere contained much less CO$_2$ than today, liquid water may have existed. Small amounts of atmospheric CO$_2$ are commonly associated with plate tectonics because of its ability to recycle carbon into the interior. It is not clear, however, whether or not early Venus possessed plate tectonics. Here, we simulate the evolution of Venus as a planet without plate tectonics, and show that weathering processes can keep the atmospheric CO$_2$ low enough to maintain liquid surface water for almost one billion years. During this time, weathering ensures that most of the CO$_2$ degassed from the interior via volcanism gets stored in the crust in the form of carbonates. Yet, part of the degassed CO$_2$ keeps accumulating in the atmosphere causing the surface temperature to rise because of the greenhouse effect. Ultimately, the liquid water on the surface evaporates and weathering stops. As soon as this happens, the crust becomes rapidly depleted in carbonates, thereby building up Venus’ CO$_2$-thick atmosphere that is observed today.
   

\section{Introduction}
\label{sec1}

The evolution of Venus is still in many ways a mystery. Despite its similarity to Earth in terms of size and bulk composition \cite<e.g.,>{Lecuyer2000}, the atmospheric CO$_2$ mass of Venus is larger by more than five orders of magnitude \cite{donahue1983}. Whether Venus' atmosphere was CO$_2$-rich early in its evolution, or whether - and how - it diverged from Earth's atmosphere later in its evolution is a matter of debate \cite<e.g.,>{gillmann2020,gillmann2016,lammer2018,lammer2009}.

What makes assessing Venus' early evolution very challenging is its relatively young surface. Most of Venus' surface has an age of $\approx$500 Ma \cite{turcotte1999,mckinnon1997}, or even less \cite{bottke2016}. Regarding its early history, both an Earth-like scenario with a temperate climate and liquid water \cite{grinspoon2007} as well as a hothouse have been discussed \cite<see>{taylor2018}.

A main requirement for habitability is the existence of liquid water \cite<e.g.,>{cockell2016}. An important constraint for liquid water on early Venus is a slow rotation enabling the formation of high-level reflective clouds, which increase the planetary albedo \cite{yang2014,way2016,way2020}. The preservation of liquid water on a planetary surface over an extended time interval is commonly associated with the presence of plate tectonics and an active carbonate-silicate cycle \cite{southam2015,Foley:2015,Kasting:2003,Sleep:2001}. If plate tectonics regulated Venus' climate in its early history, carbonates may have entered the mantle following a shallow temperature-depth gradient, and Venus may have been habitable over an extensive period of time \cite{way2020}. However, apart from limited surface features that have been associated with plume-induced subduction \cite{davaille2017}, there is no sign of active plate tectonics on Venus today \cite{smrekar2018}. The tectonic state of early Venus remains unclear, and in the absence of plate tectonics, an accumulation of CO$_2$ in Venus' atmosphere would clearly have reduced its habitable time span.

Besides plate tectonics, a major factor that is required for climate regulation is the existence of liquid water on the planetary surface. Liquid water enables silicate weathering, which forms the basis for Earth's long-term carbonate-silicate cycle \cite{Walker:1981,Kasting:2003,graham2020,honing2020}. In fact, the existence of liquid water is likely to be more important for climate regulation than plate tectonics: On stagnant-lid planets, freshly produced basaltic crust in the presence of water could be subject to silicate weathering \cite{Foley:2018,foley:2019,honing:2019}. Carbonate sediments would be buried by new volcanic eruptions and thereby move downwards. Since the stability of carbonates is strongly temperature-dependent, decarbonation would occur at depth, thereby releasing CO$_2$ \cite{Foley:2018,kerrick2001}. At shallow depth, migration of CO$_2$ to the surface may occur through cracks in porous media. Any CO$_2$ that does not migrate through cracks would sink further down as the crust is buried by new lava flows. However, the formation of partial melt below the lid and its rise towards the surface could also enable CO$_2$ to make it to the surface. The fraction of released CO$_2$ that makes it to the surface is difficult to estimate and likely depends on decarbonation depth, temperature, porosity, and crustal burial rate. Future work is required to constrain this fraction. In this paper, we assume that all released CO$_2$ makes its way to the surface, yielding the minimum habitable period. As shown by \citeA{honing:2019}, a stabilising feedback cycle consisting of silicate weathering, burial, and decarbonation can regulate the climate on stagnant-lid planets to some extent, although this is likely to be less efficient compared to a cycle with plate tectonics that recycle substantial amounts of carbon back into the mantle.

\begin{figure}
   \centering
   \includegraphics[width=\textwidth*3/4]{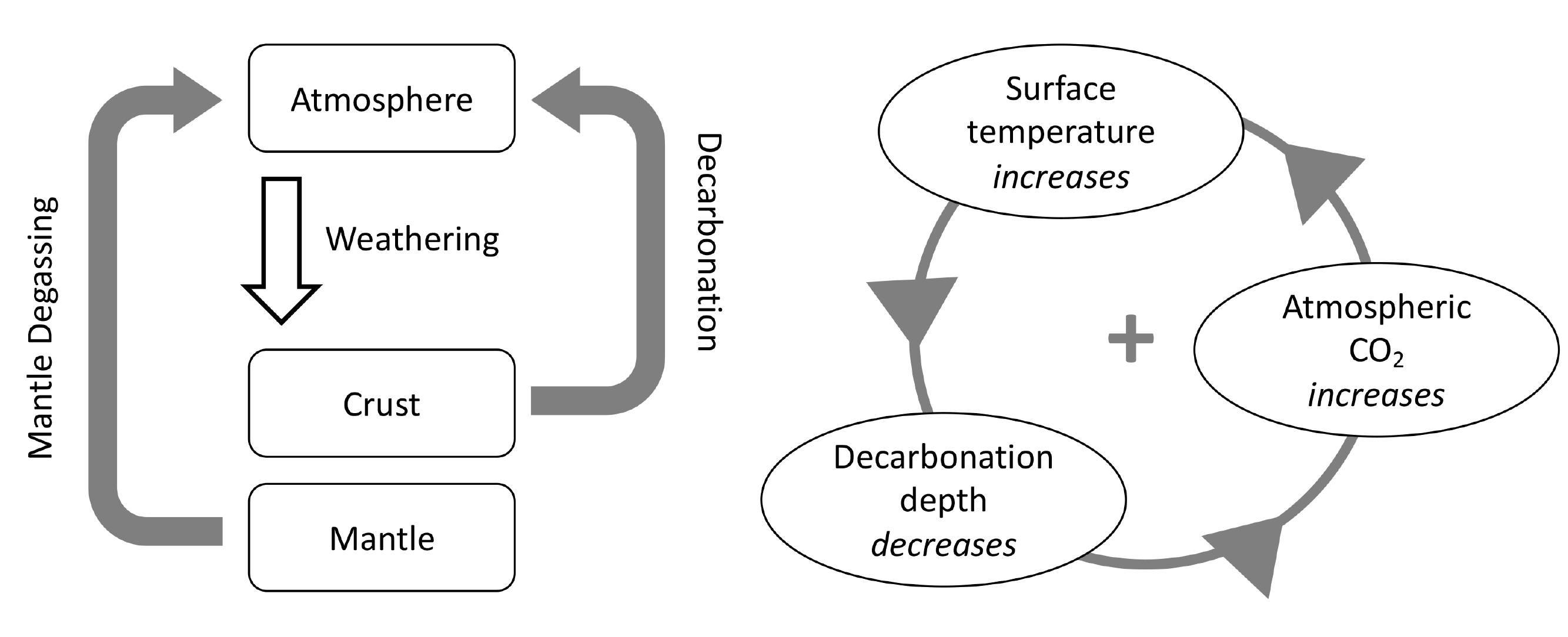}
   \caption{Left: Simplified flowchart of the carbon cycle considered in the model. Weathering only occurs if liquid water is present on the surface. Right: Illustration of the positive feedback associated with decarbonation.}
\label{fig:feedback}
\end{figure}

The carbon cycle for stagnant-lid planets outlined above (including weathering, burial, and decarbonation) features a positive feedback, a fact that has received little attention in the literature so far: The equilibrium partial pressure of CO$_2$ in the atmosphere depends upon the equilibrium between silicate weathering on the one hand, and mantle degassing plus crustal decarbonation on the other hand. The decarbonation flux depends on the rate at which carbonates reach the decarbonation depth, and therefore on the sum of (1) the rate at which carbonated crust is buried, and (2) the rate at which the decarbonation depth moves upwards. The decarbonation depth is controlled by the temperature profile in the crust, and therefore by both the mantle temperature and the surface temperature. With increasing surface temperature, the decarbonation depth moves upwards, and the decarbonation flux increases. As a result, the atmospheric CO$_2$ increases, and the surface temperature rises further (see Fig. \ref{fig:feedback}).

With increasing solar luminosity, the close proximity of Venus to the Sun makes it difficult for liquid water to be preserved on Venus' surface. With increasing surface temperature, stronger water evaporation will lead to even stronger increases in surface temperature as more water vapour is loaded into the atmosphere \cite{Driscoll:2013,kasting1988,pollack1971,ingersoll1969}. The inner boundary of the habitable zone is commonly associated with the so-called moist-greenhouse effect, where the stratosphere becomes loaded with water vapor \cite{Kasting:1993,kopparapu2013}. At the runaway greenhouse limit, the oceans evaporate entirely \cite{kopparapu2013}.

Once the inventory of liquid water has evaporated, silicate weathering would cease, so that CO$_2$ would accumulate in the atmosphere. If Venus was already in a stagnant-lid convection mode early in its history, this accumulation of CO$_2$ would have been particularly rapid and self-accelerating: as volcanism proceeds, CO$_2$ continues to be supplied to the atmosphere by both mantle degassing \textit{and} crustal decarbonation. Once silicate weathering stops, no carbon can be recycled from the atmosphere back into the crust. With the resulting rapid increase of atmospheric CO$_2$, the surface temperature then rises further whereupon crustal decarbonation speeds up (Fig. \ref{fig:feedback}b) until all available carbon has been removed from the crust and a new equilibrium state is reached. We propose that this feedback led to a sharp rise in Venus' atmospheric CO$_2$ budget and its surface temperature. In the following text, we explore the habitable period of early Venus and the subsequent self-accelerating decarbonation feedback. We also elaborate on the difference with the traditional runaway greenhouse feedback, and discuss consequences for the atmospheric evolution of stagnant-lid exoplanets.

\section{Model}
\label{sec2}
In order to assess the early evolution of Venus, we combine a thermal evolution model of the mantle with a model of carbon outgassing, weathering, burial, and metamorphic release. We assume that weathering follows the same functional dependence on crustal production and atmospheric CO$_2$ as seafloor weathering on Earth and that all carbon that is subject to metamorphic decarbonation is released back into the atmosphere. We consider CO$_2$ and water vapour as greenhouse gases and assume that weathering is active as long as liquid water is present on the surface. We test different combinations of the initial mantle temperature and mantle oxidation state and use a combination that yields Venus' observed atmospheric CO$_2$ as our reference model scenario. In the following, we provide details of all model parts.

The thermal evolution of the mantle is calculated using a parameterized model of mantle convection for stagnant-lid planets \cite{Grott:2011,morschhauser2011,Tosi:2017,godolt:2019}. The model calculates interior heat fluxes by solving the energy-conservation equations of the core, mantle, and stagnant lid, with the convective heat flux parameterized using boundary layer theory \cite{Schubert:2001}. We note that parameterized thermal evolution models commonly feature scaling laws used to calculate the heat flux which are derived from numerical models of mantle convection assuming a steady state. Although no direct comparison is available for Venus-like planets, detailed comparisons for the Moon, Mars and Mercury \cite{thiriet2019scaling} suggest that a good agreement between fully numerical calculations and parameterized models can be obtained. This indicates that our parameterized approach captures well the average characteristics of the long-term evolution of a Venus-like planet.

The mantle viscosity is calculated following \citeA{Stamenkovic:2012} as a function of temperature and pressure. The crustal production rate is calculated from the melt volume assuming solidus and liquidus of dry peridotite \cite{Katz:2003} and from the convection speed, assuming that a constant surface fraction of 1\% is covered by hot plumes in which partial melting takes place \cite{grott2010,Tosi:2017}. We note that this value is poorly constrained, however, it does not qualitatively impact our results: A smaller (larger) fraction would result in a linear decrease (increase) of the outgassing rate, and other parameter values of our reference model affecting the outgassing rate (such as the oxygen fugacity) would be adapted in a way that Venus' observed CO$_2$ is still obtained. Outgassing of CO$_2$ is based on a model of redox melting \cite{Grott:2011,hirschmann2008}. For a detailed description of the thermal evolution and outgassing model, including all relevant equations, see \citeA{Tosi:2017}.

Following \citeA{honing:2019}, weathering is calculated dependent on atmospheric CO$_2$ partial pressure ($P_{CO_2}$) as follows (see \ref{a1} for details):
\begin{linenomath}
\begin{equation}
   F_w=\frac{X_E \xi_E}{f_E}\left(\frac{\mathrm{d}M_{cr}}{\mathrm{d}t}\right)\left(\frac{P_{CO_2}}{P_{CO_2,E}}\right)^\alpha,
    \label{eq:weathersl}
\end{equation}
\end{linenomath}
where $\frac{\mathrm{d}M_{cr}}{\mathrm{d}t}$ is the crustal production rate, $P_{CO_2,E}$ is the atmospheric CO$_2$ partial pressure of present-day Earth and $\alpha\approx0.23$ is a scaling exponent \cite{Brady:1997}. The other constants in Eq. \eqref{eq:weathersl} follow from the assumption of an equilibrium between carbon degassing and recycling on present-day Earth \cite{honing:2019}: $f_E$ is the present-day Earth fraction of buried carbonates that enter the mantle, $X_E$ is the present-day Earth mid-ocean ridge CO$_2$ concentration in the melt, and $\xi_E$ is the present-day Earth fraction of seafloor weathering (see Tab. \ref{tab1} and \ref{a1}). As in \citeA{honing:2019}, we track buried carbonated crust until decarbonation occurs at a depth
\begin{linenomath}
\begin{equation}
    z_{decarb}=\frac{T_s-B}{A-\displaystyle\frac{T_m-T_s}{D_l+\delta_m }},
    \label{eq:zdecarbsl}
\end{equation}
\end{linenomath}
where $T_s$ and $T_m$ are the surface and mantle temperatures, $D_l$ and $\delta_m$ are the lid thickness and upper thermal boundary layer thickness, respectively. For the decarbonation constants $A$ and $B$ we follow \citeA{Foley:2018} and assume that the bulk metamorphic decarbonation takes place from the breakdown of dolomite (Tab. \ref{tab1}). We assume that the released CO$_2$ makes it to the surface via cracks in the crust or with uprising melt. If CO$_2$ enclosed in buried crust does not come into contact with uprising melt, it might eventually make it back into the mantle, which is a process not included in our model. For details of the model for weathering, carbonate burial, and decarbonation, we refer to \citeA{honing:2019} and \ref{a1}. 

We consider water vapour and CO$_2$ as greenhouse gases. In order to obtain a first order estimate of the habitability of early Venus, we adopt a simple two-stream radiative gray atmospheric box model to calculate greenhouse heating \cite{catling2017}. This assumes that the atmosphere is transparent to visible and other shortwave radiation, but can absorb preferentially in the infrared. This absorption is assumed to be wavelength-independent. Thereby, we neglect absorption of incoming shortwave radiation in the UV and visible by e.g. hazes and ozone. Despite these simplifications, the model reasonably well reproduces surface temperatures during Venus' early evolution as obtained by a comprehensive global climate model, as will be shown later (see Section \ref{fig:water-albedo}).

The surface temperature can be expressed in terms of the slanted optical depth $\tau$ of the atmosphere and the equilibrium temperature $T_e$:
\begin{linenomath}
\begin{equation}
    T_s^4 = T_e^4 \left( 1 + \frac{3 \tau}{4}\right) \quad \text{with} \hspace{0.6em} T_e^4 = \frac{(1-A) S_\odot}{4 \sigma},
    \label{eq:surft}
\end{equation}
\end{linenomath}
where $S_\odot$ is the incident insolation at the top of the atmosphere, $\sigma$ is the Stefan-Boltzmann constant, and $A$ is the planetary albedo. As a reference model, we mimic reflective clouds from Venus' slow rotation by adopting a planetary albedo of 0.6 \cite<consistent with>{way2020}, but also test the influence of other values for $A$. The net incident insolation is assumed to increase by a factor of 1.4 in 4.5 Gyr \cite{Gough:1981}.

We note that Eq. \ref{eq:surft} calculates the ground temperature, which is then used to calculate surface water evaporation. Equilibrium between the ground temperature and the atmospheric temperature will be reached quickly via convection, forming an adiabatic temperature profile \cite{pierrehumbert2010}. The difference between the ground temperature and near-surface atmospheric temperatures \cite<commonly defined as the temperature at 2 m, see e.g.>{deque2005} is up to $\approx$5 K \cite<e.g.,>{zeng1998}. Following \citeA{abe1985} and \citeA{pujol2003}, the optical depth of the atmosphere in the infrared is given by:
\begin{linenomath}
\begin{equation}
    \tau = \sum_i{\tau_i} = \sum_i{\frac{3 K'_i P_i}{2 g}},
\end{equation}
\end{linenomath}
where $g$ is the gravitational acceleration, $P_i$ is the partial pressure of a given atmosphere species $i$, and $K'_i$ is the absorption coefficient corresponding to this pressure. $K'_i$ can be expressed using the absorption coefficient $K_{0,i}$ at standard atmospheric pressure $P_0$
\begin{linenomath}
\begin{equation}
    K'_i = \left( \frac{K_{0,i} g}{3 P_0}\right)^{1/2},
\end{equation}
\end{linenomath}
where $K_{0,w}$ and $K_{0,c}$ are the absorption coefficients of H$_2$O and CO$_2$, respectively, as given in Tab. \ref{tab1}.

Since the hydrological cycle is not well constrained on early Venus, we make the rather straightforward assumption in our model that the atmosphere is always fully saturated in water vapour. On modern Earth, the global mean surface relative humidity is 78\% \cite{manabe1967} and large regions in the Tropics remain close to saturation. Therefore, our approach is conservative in this respect, and may underestimate the habitable period of early Venus. We calculate the saturation pressure with a vapor pressure curve from \citeA{alduchov1997}. Water in the atmosphere and the ocean are assumed to be in equilibrium, with the ocean acting as a buffer to supply water to the atmosphere as the surface temperature and pressure start to increase. As soon as all surface water is evaporated, we set the weathering rate (Eq. \ref{eq:weathersl}) to be equal to zero.

As a simplification, we neglect water outgassing from the mantle which we assume to be dry. Due to the high viscosity of a dry mantle, our successful model scenarios (that are model runs that yield outgassed CO$_2$ abundances that fit present-day observations) have higher initial mantle temperatures; otherwise our results will not be affected in any significant way (discussed in Section \ref{sec4c}). We test initial surface ocean masses of $10^{18}$ kg ($\approx$ 2.2 m Water Equivalent Layer, WEL) and $10^{19}$ kg ($\approx$ 22 m WEL), which correspond to a fraction of $\approx0.1\%$ and $\approx1\%$ of Earth's oceans and are within the bounds provided by Pioneer Venus in-situ D/H measurements \cite{Donahue1982,DonahueRussell1997}. We note that \citeA{persson2020} extrapolated the present-day escape rate of oxygen ions backwards in time (assuming present-day atmospheric composition and structure) and their results suggested that this mechanism could lead to a modest total escape of up to about 0.6 m WEL. However, additional water sink mechanisms have been proposed (discussed in Section \ref{sec4o}). 

The initial atmosphere in our model is composed of a background surface pressure of 1 bar N$_2$, which has been assumed for early Venus \cite{way2020}, for stagnant-lid planets in general \cite{Tosi:2017}, and for other habitability studies \cite{Kasting:1993}. This value is also similar to that found on Earth today and to what has been used for climate models of the Archean Earth \cite<e.g.,>{charnay2013}. Raindrop fossils indicate a surface atmospheric pressure 2.7 Gyr ago smaller than today, although this value likely fluctuated significantly \cite{som2012air,som2016earth}. All boundary parameters and initial conditions of the atmosphere and carbon cycle model are summarised in Tab. \ref{tab1}. Tab. \ref{tab2} summarises parameter values used in the thermal evolution model (see \citeNP{Tosi:2017} for a full description).

\begin{table}
\caption[]{Parameter values and initial conditions used in the atmosphere and carbon cycle model. $^{[1]}$ from \citeA{Sleep:2001} $^{[2]}$ from \citeA{Foley:2018} $^{[3]}$ from \citeA{Foley:2015} $^{[4]}$ from \citeA{Burley:2015} $^{[5]}$ from \citeA{pujol2003}}
\label{tab1}
\centering
\begin{tabular}{l l l}
\hline
Symbol & Description & Value\\
\hline
$A$ & Planetary albedo (reference) & 0.6 \\
$D$ & Orbital distance (reference) & 0.72 AU\\
$\alpha$ & Weathering exponent & 0.23 $^{[1]}$\\
$A $   & Decarbonation constant & 3.125 $\times 10^{-3}$  K $^{[2]}$\\
$B $   & Decarbonation constant & 8.355 $\times 10^{2}$  K m $^{-1}$ $^{[2]}$\\
$M_o(0)$ & Initial water ocean mass & $10^{18}$ kg, $10^{19}$ kg\\
$f_E$ & Earth's fraction of buried carbonates that enter the mantle & 0.45\\
$\xi_E$ & Earth's fraction of seafloor weathering & 0.1 $^{[3]}$\\
$X_E$ & Earth's mid-ocean ridge concentration of CO$_2$ in the melt & 125 ppm $^{[4]}$\\
$K_{0,w}$ & Absorption coefficient water vapour & 0.01 m$^2$ kg$^{-1}$ $^{[5]}$\\
$K_{0,c}$ & Absorption coefficient CO$_2$ & 0.05 m$^2$ kg$^{-1}$ $^{[5]}$\\
\hline
\end{tabular}
\end{table}

\begin{table}
\caption[]{Parameter values used in the thermal evolution model described in \citeA{Tosi:2017} with a pressure- and temperature dependent viscosity as described in \citeA{Stamenkovic:2012} $^{[1]}$.}
\label{tab2}
\centering
\begin{tabular}{l l l}
\hline
Symbol & Description & Value\\
\hline
$R_p$ & Planet Radius & 6050 km\\
$R_c$ & Core Radius & 3110 km\\
$g$ & Surface gravity & 9.0 m s$^{-2}$\\
$T_m(0)$ & Initial mantle temperature & 1700 K to 2000 K\\
$\Delta f_{O_2}$ & Mantle oxygen fugacity & IW-1 to IW+1\\
$D_{lid}(0)$ & Initial lid thickness & 50 km\\
$T_c(0)$ & Initial core temperature & 2760 K\\
$Ra_{crit}$ & Critical Rayleigh Number & 450\\
$k_m$ & Thermal conductivity (mantle) & 4 W (m K)$^{-1}$\\
$k_{cr}$ & Thermal conductivity (crust) & 3 W (m K)$^{-1}$\\
$c_c$ & Heat capacity (core) & 800 J K$^{-1}$\\
$c_m$ & Heat capacity (mantle) & 1100 J K$^{-1}$\\
$E$ & Activation energy & $3\times10^{5}$ J mol$^{-1}$ $^{[1]}$\\
$V$ & Activation volume & $2.5\times10^{-6}$ m$^3$ mol$^{-1}$ $^{[1]}$\\
$\eta_{ref}$ & Reference viscosity & $10^{21}$ Pa s $^{[1]}$\\
$T_{ref}$ & Reference temperature & 1600 K $^{[1]}$\\
\hline
\end{tabular}
\end{table}

\section{Results}
\label{sec3}
\subsection{Early evolution of Venus}
\label{sec3a}

\begin{figure}
   \centering
   \includegraphics[width=\textwidth]{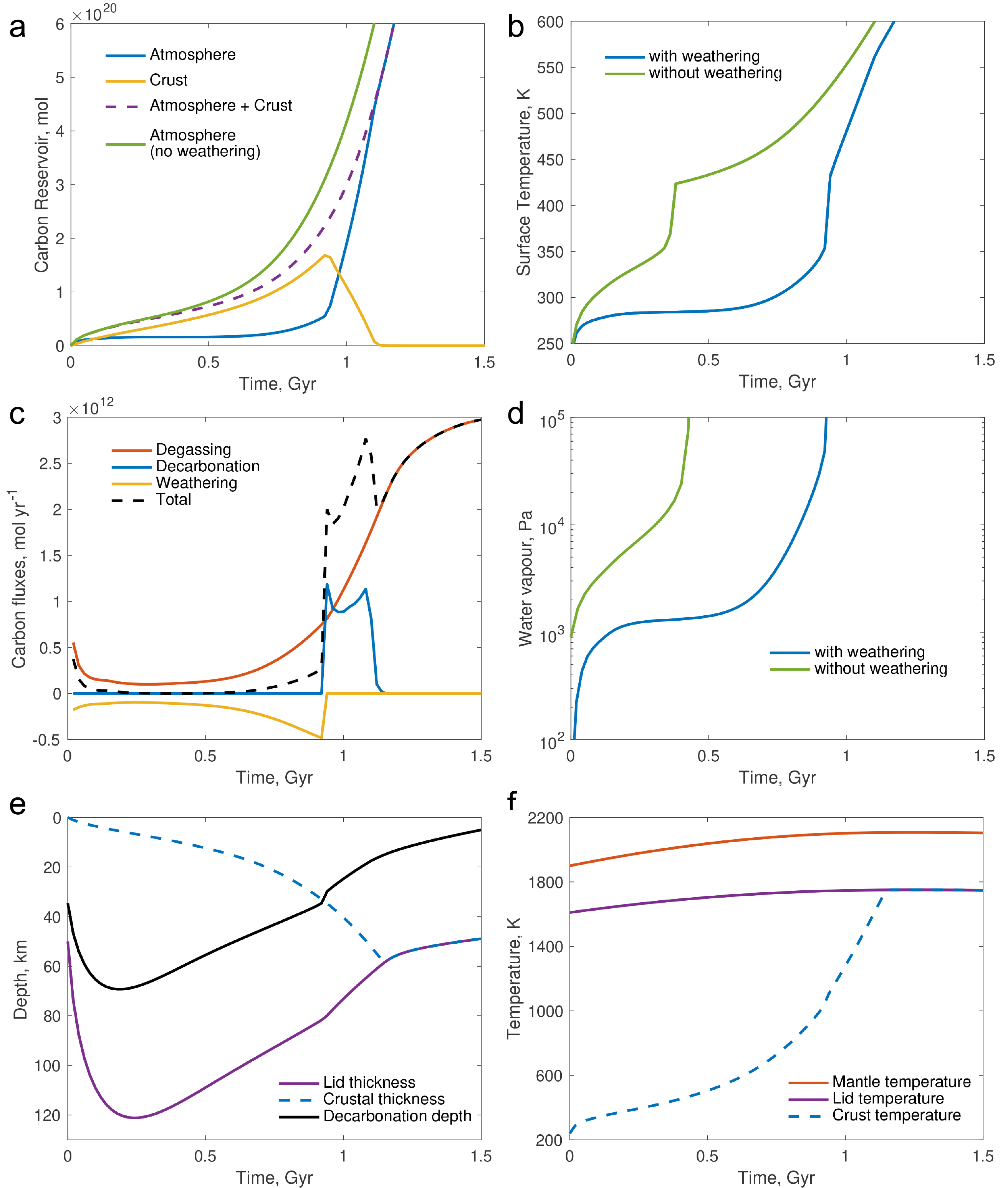}
   \caption{Early evolution of a Venus-like stagnant-lid planet at a solar distance of 0.72 AU, with an assumed planetary albedo of 0.6, a mantle oxygen fugacity equal to the iron-w\"ustite buffer, and an initial mantle temperature of 1900 K. (a) Distribution of carbon between atmosphere (blue) and crust (yellow) from a model scenario that considers weathering; the dashed purple line depicts the sum of both reservoirs. The green line depicts the total degassed CO$_2$ for a model scenario that neglects weathering. (b) Surface temperature for models with and without weathering. (c) Carbon fluxes for the scenario that considers weathering. (d) Water vapour for scenarios with and without weathering. (e) Lid and crustal thicknesses and decarbonation depth for the scenario that considers weathering. (f) Mantle, lid, and crustal base temperatures for a scenario with weathering.}
\label{fig:reference}
\end{figure}

Venus' initial mantle temperature and mantle oxidation state are not well constrained. From magma ocean solidification, initial mantle temperatures of up to 2000 K have been considered \cite{foley2014}. Since rock samples delivered from Venus have not yet been identified on Earth, the mantle oxidation state remains poorly known and may lie between reduced \cite<as on the Moon and Mars,>{Wadhwa:2008} and oxidized (as on Earth). Surface measurements by the Venera 13, 14, and Vega 2 missions suggest that the mean FeO/MnO ratio, which strongly depends on the oxidation state, likely lies between the values of Earth and Mars \cite{schaefer2017}. Fig. \ref{fig:reference} depicts results of our reference model, for which we chose an initial mantle temperature of 1900 K and an oxygen fugacity equal to the iron-w\"ustite (IW) buffer. The motivation of the choice of this parameter combination will be addressed in Section \ref{sec3b}, where we will also test other combinations of parameter values. All model runs start with an initially habitable scenario using an atmosphere without any CO$_2$, but initial abundances of CO$_2$ will be tested in Section \ref{sec4i}. Time 0 refers to the solidification of the magma ocean. In Fig. \ref{fig:reference}, panel (a) shows the early evolution of carbon in Venus' atmosphere and crust, (b) the corresponding surface temperature, (c) the carbon fluxes, (d) the water vapour, and (e) and (f) various interior evolution parameters. The green curves in Fig. \ref{fig:reference} depict the evolution of the atmospheric CO$_2$ (Fig. \ref{fig:reference}a), surface temperature (Fig. \ref{fig:reference}b), and water vapour (Fig. \ref{fig:reference}d) in a scenario where weathering is neglected so that all degassed CO$_2$ accumulates in the atmosphere. In contrast, all other curves in Fig. \ref{fig:reference} correspond to a scenario where weathering, carbonate burial and decarbonation are accounted for. The yellow curve in Fig. \ref{fig:reference}a depicts the evolution of the crustal carbon reservoir, which technically also includes a minor part of volatile CO$_2$ in the crustal matrix, introduced in order to ensure numerical stability \cite{honing:2019}. Note that the total amount of degassed CO$_2$ in the scenario that neglects weathering (green) is larger than in the scenario that includes weathering (purple dashed), since the runaway greenhouse sets in earlier, and the corresponding higher surface temperature causes a higher degassing rate during this time interval.

Fig. \ref{fig:reference} shows that including weathering considerably extends the period of time for stagnant-lid planets to remain habitable. With ongoing mantle degassing and weathering, the crustal reservoir becomes enriched in carbon whereas the atmospheric CO$_2$ only slowly increases with time during the first 900 Myr. With ongoing volcanism, buried carbonate sediments move downwards, and thereby heat up. At $\approx$900 Myr, buried carbonates become unstable. Since we assume that all CO$_2$ released by decarbonation makes it via cracks or with uprising melt to the surface, atmospheric CO$_2$ increases, and thereby the surface temperature. Liquid water evaporates and weathering stops, while the crust continues to release its carbon into the atmosphere - the planet becomes uninhabitable. However, in a scenario without weathering, the oceans evaporate already at $\approx$400 Myr. Therefore, introducing the weathering mechanism extends the potential habitable time span of a stagnant-lid Venus by $\approx$500 Myr.

It is interesting to note that in Fig. \ref{fig:reference}a the slope of the blue curve is steeper than that of the green curve at the respective time when the runaway greenhouse is triggered. This is because mantle degassing is accompanied by decarbonation while weathering has ceased. Note also that the exact value of the surface temperature directly after the runaway greenhouse (here: $\approx$430 K) follows from the (arbitrary) choice of the surface water budget. Our results however suggest that the point in time at which a planet becomes uninhabitable is not sensitive to the initial water ocean budget (shown below).

The decarbonation flux curve in Fig. \ref{fig:reference}c has several inflection points. At $\approx$ 0.92 Gyr, carbonated crust reaches the decarbonation depth (compare panel e) where carbonates become unstable. As a result, the decarbonation flux sharply increases and thereby the surface temperature, such that the surface water evaporates and weathering stops. The accompanying sharp rise of the surface temperature results in a decrease of the decarbonation depth, so that the decarbonation flux becomes particularly high. After this event, the decarbonation flux qualitatively follows the weathering flux with some delay ($\approx$0.9 Gyr), since the model tracks carbonated crust on its way to the decarbonation depth. However, since the decarbonation depth steadily decreases with time, variations in the decarbonation flux occur on a shorter timescale compared to variations in the weathering flux. In this example, variations of the decarbonation flux occur between $\approx$0.9 and $\approx$1.12 Gyr while variations of the weathering flux occur between 0 and $\approx$0.9 Gyr. At $\approx$1.12 Gyr, all carbonates have become unstable, and the decarbonation flux approaches zero. It does not however instantaneously approach zero, because of the volatile CO$_2$ reservoir introduced into the crustal matrix, which releases a constant fraction to the atmosphere at each timestep.

As long as the combined outgassing flux does not vary significantly with time and weathering is active, an equilibrium between outgassing and weathering is expected, since weathering is controlled by atmospheric CO$_2$ \cite<e.g.,>{berner1997need}. In Fig. \ref{fig:reference}c (black dashed line), we however observe that the total carbon flux is not always equal to zero and therefore the system is not always in steady state. In the early evolution (0.1 Gyr), equilibrium between degassing and weathering is not reached and CO$_2$ in the atmosphere builds up. Later in the evolution (0.6 Gyr), degassing substantially increases, which is a result of the assumed dry mantle of high viscosity and the thereby increasing mantle temperature (compare Fig. \ref{fig:reference}f). Due to the increasing degassing rate, the atmospheric CO$_2$ subsequently increases. As a result, the weathering rate also increases, but with some delay since some CO$_2$ has accumulated in the atmosphere. Once weathering has ceased, atmospheric CO$_2$ rapidly increases.

\begin{figure}
   \centering
   \includegraphics[width=\textwidth]{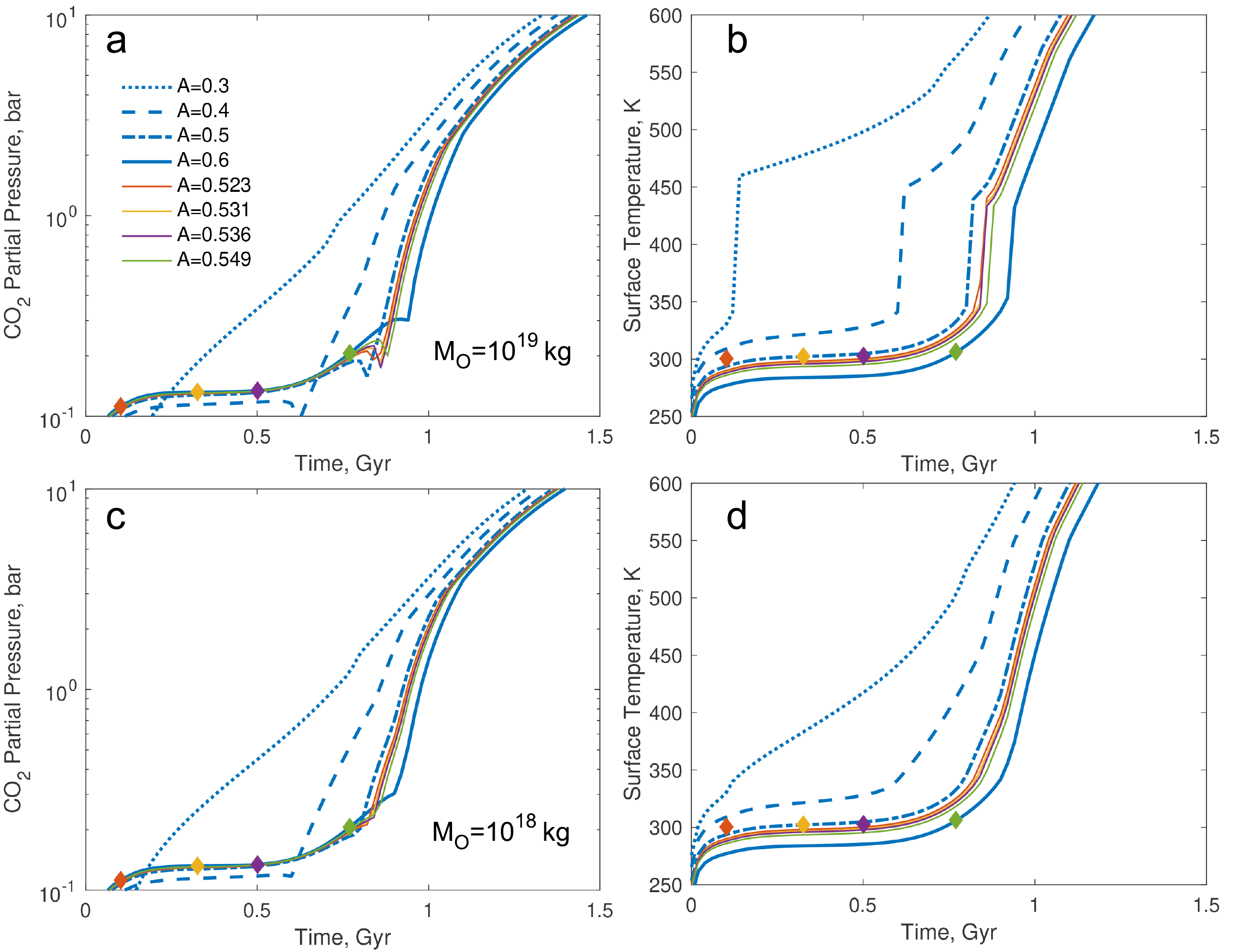}
   \caption{Comparison between an initial ocean mass of  $10^{19}$ kg (top) and $10^{18}$ kg (bottom) and different planetary albedos A. The diamonds in (a) and (c) represent combinations of incident insolation and atmospheric CO$_2$ obtained from our reference model, which were used as an input for the 3D global climate model assuming a water equivalent layer of 10m \cite{way2020}. The resulting average surface temperatures are plotted as diamonds in (b) and (d), and the resulting albedos are used as inputs for additional runs of our parameterized atmosphere-interior model, which are plotted as red, yellow, purple, and green curves. See text for further explanation.}
\label{fig:water-albedo}
\end{figure}

In Fig. \ref{fig:water-albedo}, we compare the early evolution of the atmospheric CO$_2$ and surface temperature using initial ocean masses of $10^{19}$ kg (top) and $10^{18}$ kg (bottom). As expected, the effect of the initial ocean mass on the early evolution before the runaway greenhouse is negligible, whereas the temperature increase during the runaway greenhouse is more pronounced with a larger ocean reservoir. The slight decrease of CO$_2$ partial pressure in Fig. \ref{fig:water-albedo}a before the sharp inflection is a result of the rapid water evaporation, which reduces the atmospheric volume fraction of CO$_2$ to a greater extent than it increases the total mass of the atmosphere. In Fig. \ref{fig:water-albedo}c however, the water mass added to the atmosphere is one order of magnitude smaller, hence this effect is not apparent here.

In addition, Fig. \ref{fig:water-albedo} depicts the effect of the planetary albedo. With an Earth-like albedo of 0.3 (dotted curves), the runaway greenhouse is triggered directly in the early stages as soon as a relatively small amount of CO$_2$ is released into the atmosphere. In particular an initial atmospheric CO$_2$ reservoir would cause a direct transition from the magma ocean phase to a runaway greenhouse. However, slow rotation in Venus' early evolution could have favoured the formation of a global, reflective cloud layer, increasing the planetary albedo to a value between 0.5 and 0.6 \cite{way2016,way2020}. Fig. \ref{fig:water-albedo} shows that this can lead to an extended early habitable period, even in a stagnant-lid regime. In addition to the point in time where the runaway greenhouse effect is triggered, the planetary albedo influences the surface temperature before and after this event.

In order to compare our results for early Venus with output from comprehensive 3D climate models, we use the 3D General Circulation Model (GCM) ROCKE-3D \cite{way2020,way2017}. We set the atmospheric composition to 1 bar N$_2$ and test four parameter combinations of atmospheric CO$_2$ and incident insolation that fit our reference model scenario during Venus' early evolution: (i) 0.112 bar surface CO$_2$ and an insolation of 1886.1 W m$^{-2}$, (ii) 0.132 bar CO$_2$ and 1913.6 W m$^{-2}$, (iii) 0.134 bar CO$_2$ and 1935.4 W m$^{-2}$, (iv) 0.206 bar CO$_2$ and 1970.1 W m$^{-2}$. In Fig. \ref{fig:water-albedo}a and c, we plot these points as diamonds. From the GCM, we obtain the following planetary albedos and average surface temperatures (averaged over the last 50 years of each run after they have reached radiative equilibrium): (i) an albedo of 0.523 and surface temperature of 300.4 K, (ii) 0.531 and 302.3 K, (iii) 0.536 and 302.7 K, (iv) 0.549 and 306.4 K. Using these planetary albedos, we then re-run our parameterized interior-atmosphere model and plot the results as (i) red, (ii) yellow, (iii) purple, and (iv) green curves in Fig. \ref{fig:water-albedo}. In Fig. \ref{fig:water-albedo}b and d, we plot the average surface temperatures obtained from the GCM model as diamonds. We find that the results from our parameterized interior-atmosphere model during the early evolution are in good agreement with the GCM model, particularly on the plateau region in atmospheric CO$_2$ and surface temperature around $\approx$0.5 Gyr if the respective albedos derived from the GCM are used. All four control points confirm our main finding of an early habitable stagnant-lid Venus.

\subsection{Influence of mantle oxidation state and initial mantle temperature}
\label{sec3b}

\begin{figure}
   \centering
   \includegraphics[width=\textwidth]{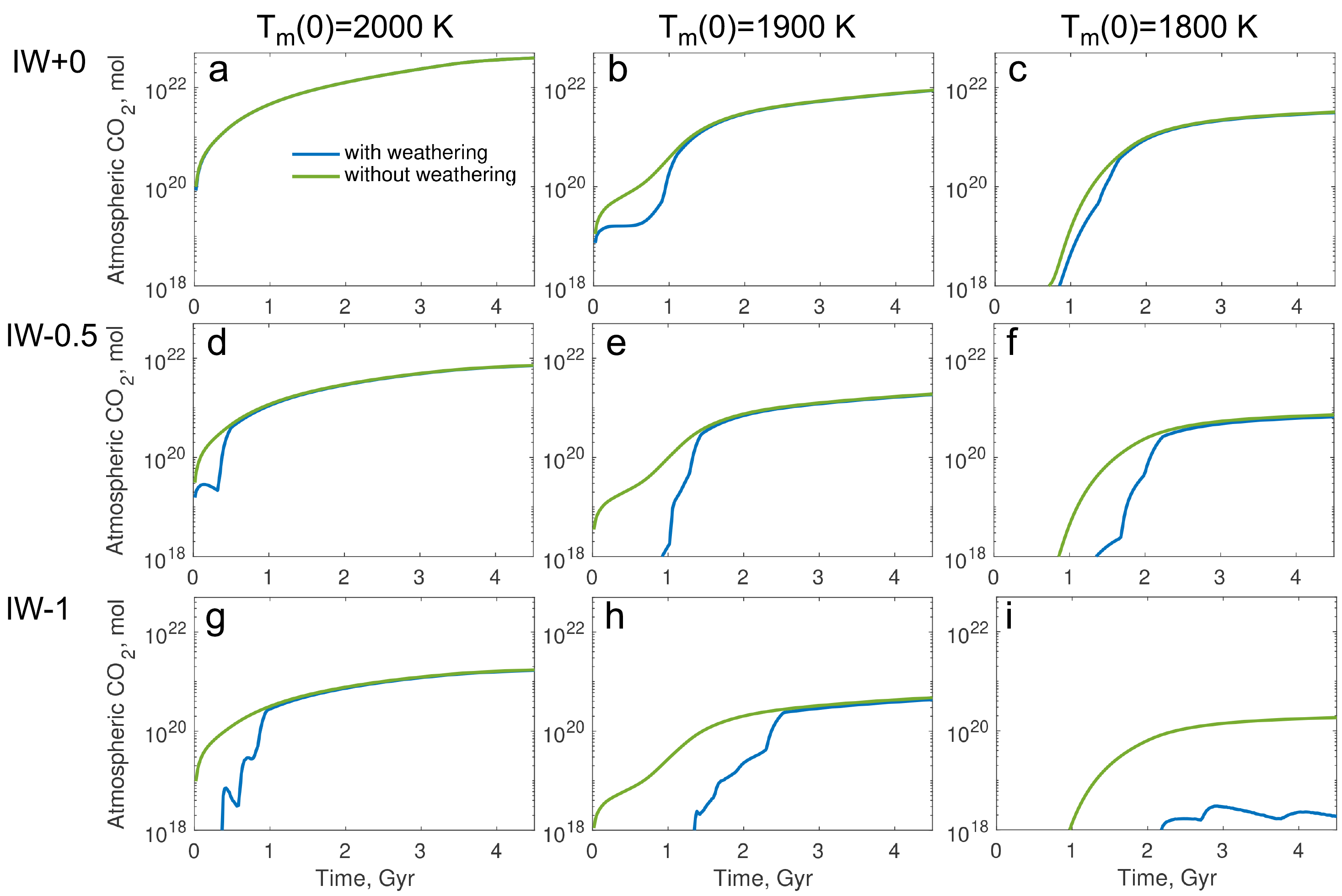}
   \caption{Evolution of the atmospheric CO$_2$ for a stagnant-lid planet at 0.72 AU for model scenarios with (blue) and without (green) weathering and different oxygen fugacities (top row: IW, middle row: IW-0.5, bottom row: IW-1) and initial mantle temperatures (left column: 2000 K, middle column: 1900 K, right column: 1800 K).}
\label{fig:parameter}
\end{figure}

The oxygen fugacity and the initial mantle temperature are major factors that could strongly impact the atmospheric evolution of stagnant-lid planets for models with \cite{honing:2019} and without weathering \cite{ortenzi2020,Tosi:2017,godolt:2019}, in particular in combination with the surface pressure \cite{gaillard2014}. Unfortunately, these parameter values are poorly known for Venus. However, Venus' present-day atmosphere contains $\approx10^{22}$ mol CO$_2$, which may be used as a boundary condition. Continuous volcanic degassing in a stagnant-lid regime is in agreement with the observed argon mass in Venus' atmosphere \cite{orourke2015}, although we note that the use of argon to constrain the degassing history of terrestrial planets is not straightforward and that the solutions are not unique \cite{watson2007,way2020}.

Fig. \ref{fig:parameter} shows the evolution of atmospheric CO$_2$ for various parameter combinations of the initial mantle temperature (2000 K, 1900 K, 1800 K) and oxygen fugacity (iron-w\"ustite buffer IW, half a log unit below, IW-0.5, and one log unit below, IW-1). The highest initial mantle temperature that we test (2000 K) is motivated by considerations of magma ocean crystallisation \cite{foley2014}. The solar distance is 0.72 AU and the model accounts for the solar flux evolution. Green curves again show model scenarios that neglect weathering whereas blue curves show scenarios that consider weathering, carbonate burial, and decarbonation.

Fig. \ref{fig:parameter} suggests that a high initial mantle temperature as well as a high oxygen fugacity both favour a high atmospheric CO$_2$ amount throughout the entire evolution due to a high mantle degassing rate. This is the case for the model scenarios with and without weathering. The point in time where the blue curve approaches the green curve can be interpreted as the onset of the runaway greenhouse of the model scenario with weathering: As soon as both curves converge, the crust has lost all its carbon, which is a direct consequence of the halt to weathering due to a lack of liquid water. It therefore becomes apparent that the weathering feedback becomes particularly important for maintaining habitability if the oxygen fugacity and the initial mantle temperature are relatively low. Fig. \ref{fig:parameter}i suggests that for a small degassing rate throughout the entire evolution, weathering can keep the atmospheric CO$_2$ sufficiently low to avoid a runaway. This result is consistent with \citeA{way2020}, who find that for low atmospheric CO$_2$, Venus could have remained habitable until today. The three local maxima in Fig. \ref{fig:parameter}i reflect changes of the decarbonation flux: As soon as crust with a higher carbon concentration than that of the previous timestep reaches decarbonation depth, the decarbonation flux increases, and thereby the weathering rate and the carbon concentration of freshly formed crust. In combination with a steadily increasing decarbonation depth, these events result in local maxima \cite<see also Fig. 2 of >{honing:2019}. We note, however, that the total amount of degassed CO$_2$ (green curve) at 4.5 Gyr is almost two orders of magnitude lower than that observed for Venus today, indicating that the scenario depicted in Fig. \ref{fig:parameter}i is not representative of Venus' evolution.

We note the different shapes of the blue curves in Fig. \ref{fig:parameter}. For example, the blue curve of the top centre model (IW+0, 1900 K, Fig. \ref{fig:parameter}b) is approximately constant until 900 Myr, which is the point in time where the runaway greenhouse is triggered (compare with Fig. \ref{fig:reference}). In contrast, if the initial mantle temperature is higher (for example 2000 K, IW-0.5), the atmospheric CO$_2$ experiences a local maximum. This is because a hot mantle rapidly cools down in the early evolution. A different behaviour can be seen for cases where the initial mantle temperature is lower (for example IW-0.5, 1800 K, Fig. \ref{fig:parameter}f). Here, the atmospheric CO$_2$ steadily increases.

It is interesting to note that a steadily increasing atmospheric CO$_2$ is contrary to what is assumed for Earth's atmospheric evolution \cite<e.g.,>{Sleep:2001}. The reason for this is twofold: First, plate tectonics planets lack a thick, insulating lid, and therefore cool down relatively quickly \cite<e.g.,>{Schubert:2001,Honing:2016}. As a result, the degassing rate on planets with plate tectonics decreases with time, for example due to a decreasing plate speed \cite{oosterloo2021} or an increasing fraction of subduction zones that avoid decarbonation \cite{honing:2019}. Second, we assume that weathering on stagnant lid planets is a function of CO$_2$ only, whereas continental weathering on Earth is a function of both CO$_2$ and temperature. With increasing incident insolation, a smaller amount of CO$_2$ is therefore required for continental weathering to balance degassing and the atmospheric CO$_2$ partial pressure decreases with time.

Comparing the present-day atmospheric CO$_2$ (t=4.5 Gyr) of each model run with the present-day atmospheric CO$_2$ of Venus ($\approx10^{22}$ mol) provides a first-order estimate of a reasonable initial mantle temperature and oxygen fugacity of Venus. We find that an oxygen fugacity equal to the iron-w\"ustite buffer and an initial mantle temperature of 1900 K fits the present-day observation well. Therefore, we chose this parameter combination as a reference case, which has been explored in more detail in Fig. \ref{fig:reference}. Note however, that this solution is not unique; a higher initial mantle temperature combined with a smaller oxygen fugacity, or vice versa, could also fit the present-day observation of Venus' atmospheric CO$_2$ (see e.g. Fig. \ref{fig:parameter}d).

\subsection{Bimodal distribution of atmospheric CO$_2$}
\label{sec3c}
While the atmospheric CO$_2$ on planets with active weathering remains low, a halt to weathering would naturally lead to increasing atmospheric CO$_2$ with time, independent of the tectonic state of the planet. Since planets with plate tectonics are anticipated to have higher degassing rates than stagnant-lid planets, CO$_2$ accumulation through mantle degassing would be particularly rapid if plate tectonics keeps operating. In addition, the surface temperature after a runaway greenhouse could become so high that carbonate sediments on the surface become unstable, which would result in a particularly strong increase in atmospheric CO$_2$ \cite{graham2020}. As a consequence, the atmospheric CO$_2$ would show a bimodal distribution with low abundances as long as weathering is active and higher abundances thereafter.

On stagnant-lid planets, the mechanisms described above would likely operate less efficiently due to smaller degassing rates. Here, however, an additional mechanism would come into play, leading to a bimodal distribution in atmospheric CO$_2$. Due to the lack of subduction zones, degassed CO$_2$ would not be recycled back into the mantle. Instead, as long as weathering is active, the crust would be gradually enriched in carbonates (see Fig. \ref{fig:feedback}, left, and Fig. \ref{fig:reference}a), while atmospheric CO$_2$ would remain low. As soon as weathering ceases, ongoing crustal burial and decarbonation would rapidly deplete the crust in carbonates. In Fig. \ref{fig:reference}a we observe an increase of atmospheric CO$_2$ by approximately one order of magnitude within 100 Myr.

\begin{figure}
   \centering
   \includegraphics[width=\textwidth*3/4]{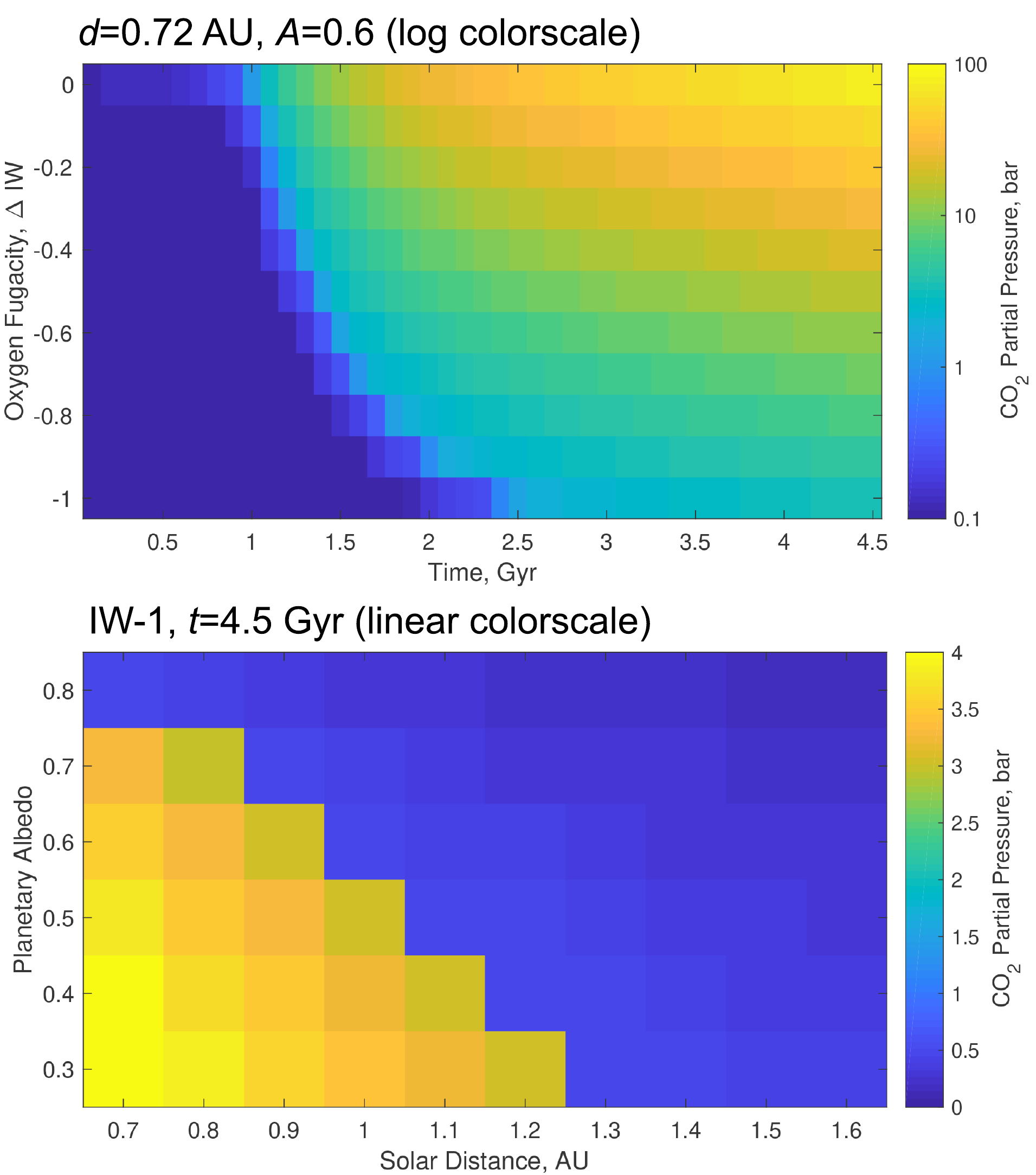}
   \caption{Top: Atmospheric CO$_2$ for a stagnant-lid planet at 0.72 AU with a planetary albedo of 0.6 as a function of oxygen fugacity and time. Bottom: Atmospheric CO$_2$ of a stagnant-lid planet at 4.5 Gyr for an oxygen fugacity of IW-1 as a function of solar distance and planetary albedo. All model runs have an initial mantle temperature of 1900 K and assume active weathering as long as liquid water is present on the surface.}
\label{fig:distance}
\end{figure}

An example of the aforementioned bimodal distribution is depicted in Fig. \ref{fig:distance}, where we show the atmospheric CO$_2$ partial pressure of a Venus-sized stagnant-lid planet as a function of oxygen fugacity and time (top) and of planetary albedo and solar distance (bottom). Note that for our rather straightforward atmospheric model it is challenging to estimate the surface temperature and thereby the transition to a runaway greenhouse for high CO$_2$ abundances. With this in mind, in Fig. \ref{fig:distance} (bottom), we chose a combination of initial mantle temperature (1900 K) and mantle oxidation state (IW-1) that results in a relatively modest amount of atmospheric CO$_2$ of up to a few bar at t=4.5 Gyr. This range is comparable to habitable zone planets considered by \citeA{Tosi:2017} and \citeA{kadoya2019}.

The atmospheric CO$_2$ dramatically varies depending on whether (green/yellow) or not (blue boxes) the planet experienced a runaway greenhouse in its history, with a difference of approximately one order of magnitude. For a planet at Venus' distance and an albedo of 0.6 (Fig. \ref{fig:distance}, top), atmospheric CO$_2$ remains in the order of 0.1 bar as long as weathering is active and increases to more than 1 bar after weathering has ceased; this transition is only marginally affected by the oxygen fugacity. For stagnant-lid planets with an oxygen fugacity of IW-1 and an age of 4.5 Gyr (Fig. \ref{fig:distance}, bottom), we obtain a bimodal distribution between 0.7 and 1.3 AU depending on the planetary albedo. The difference in atmospheric CO$_2$ between planets with and without weathering remains in the order of at least one order of magnitude.

This finding has profound implications for future exoplanet observations. If the CO$_2$ absorption band retrieval were to indicate enhanced atmospheric abundances for rocky planets lying closer to their host stars (assuming the planets do not lie too close to the star so that escape of C and O atoms becomes relevant) than for planets further away, then this might indicate that an active weathering cycle operating on planets of the latter category, even if these are in a stagnant-lid regime. A planet observed with a small atmospheric CO$_2$ budget could give an indication of liquid surface water and a habitable climate. CO$_2$ absorption may be one of the few observable spectral signatures for favourable rocky exoplanetary targets with current and near-future platforms \cite<e.g.,>[]{wunderlich2021,Fauchez2019}. For this reason the interpretation of the CO$_2$ abundances following a bimodal distribition could be critical to assess exoplanet habitability since other variables relevant for habitability such as the possible presence of additional greenhouse gases are generally more challenging to determine \cite{Fauchez2019}.

\section{Discussion}
\label{sec4}
We coupled a thermal evolution model for stagnant-lid planets with a model of weathering, carbonate burial, and decarbonation, and applied this model to early Venus and stagnant-lid exoplanets. Our main findings are twofold. First, crustal weathering could yield an extensive habitable period for early Venus even in a stagnant-lid regime, and second, stagnant-lid exoplanets may show a bimodal distribution of their atmospheric CO$_2$. Assumptions and uncertainties in the model that could affect these conclusions are discussed below.

\subsection{Limitations of the climate model}
\label{sec4l}
Since our coupled interior-atmosphere evolution model calculates reservoirs and fluxes over billions of years, it was necessary to make substantial simplifications regarding the atmospheric climate model. Instead of a radiative-convective model as is commonly used for approximating the runaway greenhouse \cite<e.g.,>{kopparapu2013,kopparapu2017}, we adopted a simple two-stream radiative grey atmospheric box model \cite{catling2017}. Assuming constant planetary albedo and opacity, results can only be seen as reliable if changes in atmospheric CO$_2$ and water vapour remain small. Nevertheless, the early evolution of Venus is simulated reasonably well with our model. In Fig. \ref{fig:water-albedo}, we show that the temperature difference obtained from our model compared with results from a 3D global climate model at 4 control points during the early habitable period is relatively small ($\approx$10 K if compared to a model run with the respective albedo obtained from the GCM, and if compared to a model run of a constant albedo of 0.6, only the first three control points show a moderately higher temperature difference of $\approx$20 K). All four control points confirm our main finding of early habitable conditions. We note that our assumptions of constant albedo and opacity do not hold for predicting the climate for the period after ocean water evaporation. However, in Fig. 4, where we plot the complete evolution over 4.5 Gyr, we only show the atmospheric reservoir. This reservoir is mainly an outcome of the interior evolution and would therefore be affected by the climate to a much lesser extent.

\subsection{Parameterisation of weathering and CO$_2$ degassing}
\label{sec4c}

On Earth, seafloor weathering is an important component of the long-term carbonate-silicate cycle \cite{Brady:1997,Krissansen-Totton:2017,krissansen-totton:2018}. Hydrothermal alteration of new basaltic crust produces alkalinity, which favours precipitation and burial of carbonates. This concept has been applied to stagnant-lid planets \cite{Foley:2018,foley:2019,honing:2019}. However, even on Earth, the dependence of the rate of seafloor weathering on atmospheric CO$_2$, temperature, and/or ocean pH is a matter of debate \cite{krissansen-totton:2018,Krissansen-Totton:2017,Coogan:2015,Sleep:2001,Brady:1997}. The effect of different parameterizations for seafloor weathering on the climate of stagnant-lid planets has been explored in \citeA{honing:2019}. These authors find that a parameterization that neglects any direct dependence of the weathering rate on atmospheric CO$_2$ inevitably causes a removal of almost the entire atmospheric reservoir as soon as the degassing rate approaches zero, which appears to be unrealistic. Similarly, using a solely temperature-dependent parameterisation for seafloor weathering, model results by \citeA{foley:2019} suggest a low abundance of atmospheric CO$_2$ and an icy surface as soon as outgassing ceases. We decided to use a direct dependence of basalt weathering on atmospheric CO$_2$ in our model, but note that additional dependencies on surface temperature or ocean pH cannot be ruled out and might affect the habitable time span of early Venus. In addition, depending on the poorly constrained topography shifts of early Venus, a thermodynamic limit to weathering \cite{graham2020,maher2014} might also be worth to be considered in follow-up work.

The assumed direct dependence of the weathering rate on the crustal production rate is fundamentally different than on Earth, where weathering proceeds mainly via continental processes, which depend on mountain building and erosion exposing fresh rock. Since we assume that on stagnant-lid planets freshly produced basaltic crust is directly subject to weathering, it inevitably follows that even if erosion takes place later in the evolution, the crust exposed as a result would already have been carbonated at the time it formed \cite<see>{honing:2019}. Therefore, we argue that on stagnant-lid planets, weathering rates are likely to depend proportionally on crustal production, even if the crust is not submerged. We however point out the importance of liquid water for weathering of freshly formed crust. If part of the surface of early Venus were emerged above sea-level, which is indeed likely to be the case for the smaller assumed initial ocean reservoir (Fig. \ref{fig:water-albedo}c and d), this would require regular rainfall.

Assuming that weathering on early Venus has a similar functional dependence on CO$_2$ as seafloor weathering on Earth, we assumed that the mineralogy of Venus is mainly basaltic. However, recent observations from the Venus Express mission indicate that the mineralogy of Tessera Terrain, a region of high topography, is more silica-rich or felsic \cite{gilmore2017}. On the one hand, weathering fluxes at a given CO$_2$ for granites may be smaller than for basalts \cite{hakim2021}, and therefore on average, the weathering exponent $\alpha$ (Eq. \ref{eq:weathersl}) might be smaller. On the other hand, for our reference model, we find that the habitable period ends as soon as decarbonation of the crust occurs, which weakens the dependence of our results upon the exact value of $\alpha$. In model scenarios where the planet remains habitable after decarbonation sets in (see Fig. \ref{fig:parameter}), a smaller value of $\alpha$ would reduce the habitable period. Altogether, a better knowledge of Venus' mineralogy and a better understanding of its crustal production history is needed in order to further improve our understanding of its early habitability. A significant step towards the determination of Venus' surface mineralogy could be taken with the recently approved VERITAS mission \cite{freeman2016veritas}.

Whether or not weathering could have extended the habitable time span of an early stagnant-lid Venus depends on its initial mantle temperature and oxygen fugacity. If the combination of both parameters exceeds a certain threshold, then rapid outgassing of CO$_2$ inhibits efficient climate regulation (see Fig. \ref{fig:parameter}). In particular an initial mantle temperature as high as 2000 K would dramatically limit the habitable period. However, the combination of an initial mantle temperature of 1900 K together with an oxygen fugacity equal to the iron-w\"ustite buffer allows for a habitable period of 900 Myr while the atmospheric CO$_2$ amount obtained at 4.5 Gyr is in agreement with present-day observations of Venus' atmospheric CO$_2$.

\begin{figure}
   \centering
   \includegraphics[width=\textwidth*3/4]{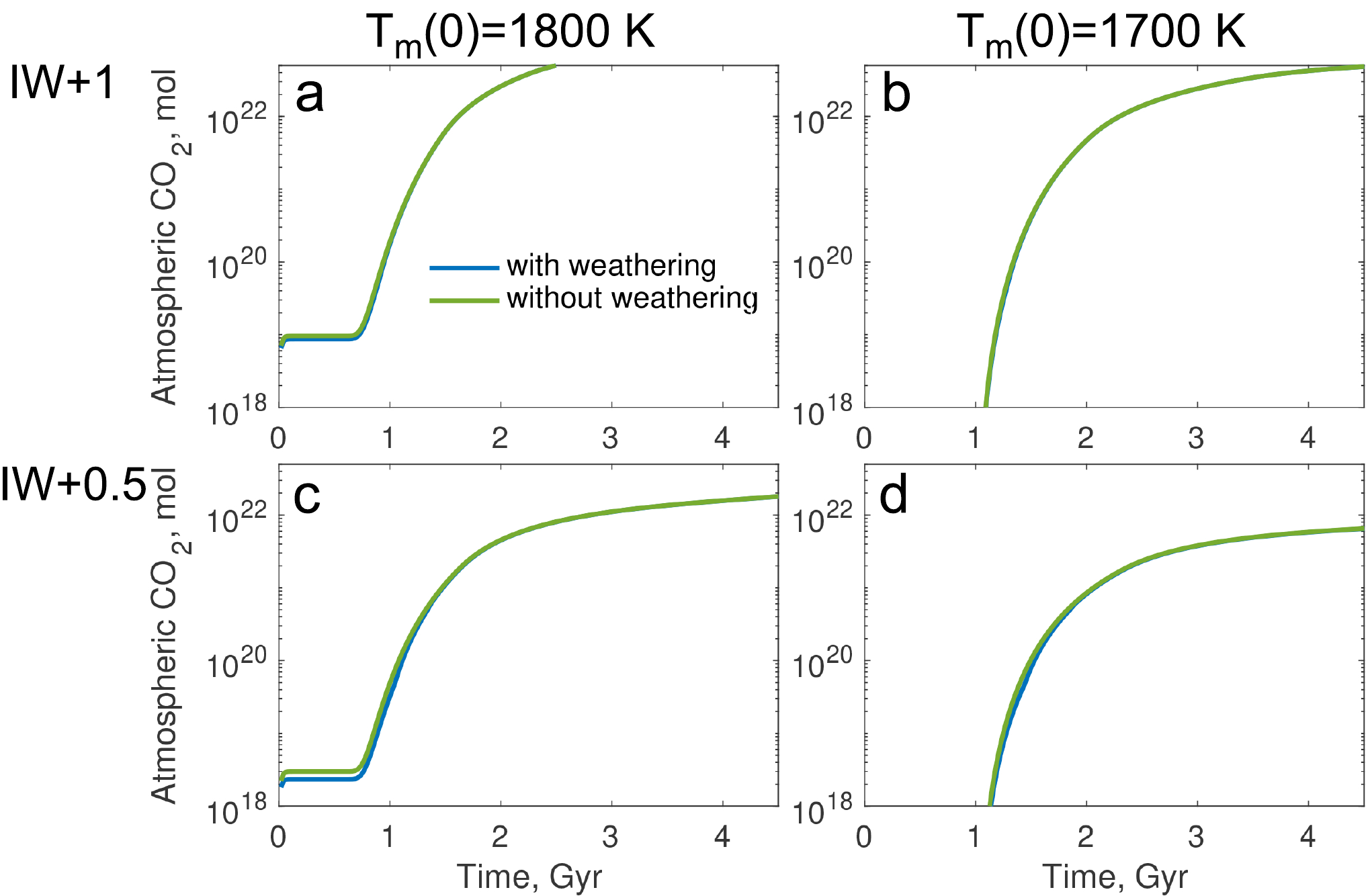}
   \caption{Evolution of atmospheric CO$_2$ for a planet as in Fig. \ref{fig:parameter} but for an initial mantle temperature of 1800 K (left column) and 1700 K (right column) in combination with an oxygen fugacity of IW+1 (top row) and IW+0.5 (bottom row).}
\label{fig:highiw}
\end{figure}

In Fig. \ref{fig:parameter}, we explored parameter combinations of an initial mantle temperature of 1800 and 2000 K and an oxygen fugacity between the iron-w\"ustite Buffer and one log-unit below it. However, since no rock samples from Venus have been identified yet, its oxidation state is unknown, and an even higher oxidation state - similar to Earth - cannot be ruled out. In Fig. \ref{fig:highiw}, we show that a combination of a smaller initial mantle temperature and a higher oxygen fugacity (for example 1700 K and IW+0.5) yields an atmospheric CO$_2$ partial pressure similar to that obtained in the reference scenario (with 1900 K and IW+0). However, the effect of weathering is negligible here, since the small initial degassing rate caused by the low temperature results in low atmospheric CO$_2$ during the first $\approx$1 Gyr in both model scenarios. Later in the evolution, the combination of a high degassing rate (from a hotter mantle) and a high solar flux triggers a runaway independent of whether or not weathering is active.

An additional simplification of our model is that it does not explicitly calculate Venus' mantle carbon inventory. Earth's carbon inventory has been estimated at $2.5\times10^{22}$ mol \cite{Sleep:2001}, but could be as high as $1\times10^{23}$ mol \cite{dasgupta2010}. Assuming that Venus' initial mantle carbon inventory is similar to Earth's, explicitly modelling Venus' mantle carbon reservoir would not significantly affect early degassing during the habitable period. In the late evolution, our simplification may overestimate degassing if mantle material previously depleted in CO$_2$ is brought to the melt region. Since our model assumes that melting occurs within plumes, which are expected to sample deep primordial mantle material, this is not necessarily the case, however. On the other hand, we also neglected external sources that could have supplied carbon to the atmosphere, for example during late accretion events \cite<see for example>{gillmann2020}, however, these sources would not dominate the planetary carbon budget \cite<for Earth, see>{mikhail2019}.

In addition to the initial mantle temperature and the oxygen fugacity, water in the mantle would affect the interior evolution and the degassing rate. First, it would reduce the viscosity, and thereby enhance the convection rate, heat flow, and CO$_2$ degassing. Second, it would reduce the melting temperature, which further enhances degassing. In order to limit the complexity of our model and to allow for a more straightforward way of studying relevant feedbacks, we neglected water in the mantle and its degassing into surface oceans. Calculating the mantle viscosity, we followed \citeA{Stamenkovic:2012}, thereby neglecting the water-dependence of viscosity. As shown by \citeA{Tosi:2017}, an initially wet mantle can significantly enhance early degassing of CO$_2$ and would thereby compete with a high initial mantle temperature in order to successfully reproduce the atmospheric CO$_2$ of present-day Venus. A wet mantle would allow for an initial mantle temperature of 1700 K or lower to yield early degassing of CO$_2$ \cite{Tosi:2017}, although the cycle of weathering, carbonate burial, and decarbonation as described above would likely operate similarly.

\subsection{Initial ocean mass}
\label{sec4o}

From analysis of large-probe neutral mass spectrometer data obtained from Pioneer Venus, early Venus is thought to have a minimum water mass of $10^{18}$ kg on its surface, which is two orders of magnitude more than its present-day amount \cite{donahue1992}. Some studies even suggest a more massive water ocean on early Venus \cite{salvador2017}, but the low present-day oxygen ion escape rate of Venus makes such a scenario challenging to justify \cite{persson2020,gillmann2020,lammer2009}. A potentially significant water sink has been suggested by \citeA{way2020}: During the global resurfacing event $\approx$ 500 Myr ago, a large volume of freshly produced basaltic crust may have been an efficient oxygen sink \cite{way2020}. We note that a global resurfacing event has been associated with a transition from plate tectonics to stagnant-lid \cite{solomatov1996stagnant}, which we do not assume in our model. However, numerical simulations show that such a resurfacing event may also emerge from a stagnant-lid convection regime \cite{noack2012}. \citeA{armann2012simulating} and \citeA{Gillmann:2014} find that simulations with an episodic overturn can more readily explain surface observations such as topography, gravity, surface age, and inferred crustal thickness, although a continuous stagnant-lid regime cannot be excluded for specific model parameter combinations. \citeA{rolf2018inferences} suggest that simulations and observations best favour at least one global overturn event with ongoing resurfacing. Altogether, continuous stagnant-lid convection with one or more resurfacing events as efficient oxygen sinks in Venus' late evolution remains a valid option.

Modelling Venus' early evolution (Fig. \ref{fig:reference}), we decided to use an initial surface ocean reservoir of $10^{19}$ kg. This value is sufficiently high to illustrate a significant runaway greenhouse effect, resulting in a surface temperature increase of $\approx$100 K (see Fig. \ref{fig:reference}b). However, since our model neglects a water sink, this high value would result in an overestimation of the surface temperature and thereby mantle melting and degassing in the late evolution. Therefore, when plotting the atmospheric CO$_2$ in Fig. \ref{fig:parameter}, we used for this case an initial surface ocean reservoir of $10^{18}$ kg. The early evolution before the runaway greenhouse and its onset is not sensitive to the assumed ocean mass (see Fig. 3).

\subsection{Initial CO$_2$-rich atmosphere}
\label{sec4i}

\begin{figure}
   \centering
   \includegraphics[width=\textwidth]{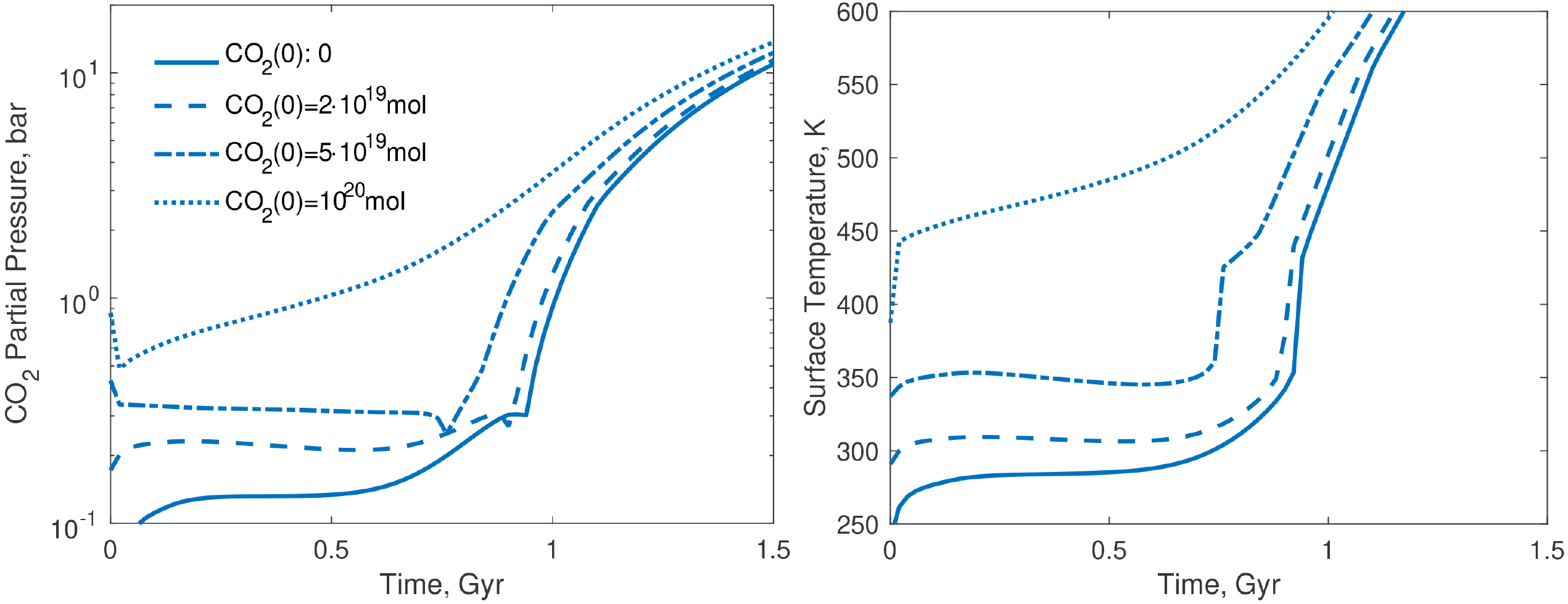}
   \caption{Early evolution with initial CO$_2$ abundances between 0 and 1$\times10^{20}$ mol, other parameter values as in Fig. \ref{fig:reference}.}
\label{fig:initial}
\end{figure}

For the sake of simplicity, our reference model assumes an initial atmosphere only composed a background-pressure of 1 bar N$_2$. However, magma ocean solidification might degas a substantial amount of CO$_2$ into the atmosphere \cite{elkins2008,lebrun2013,salvador2017,nikolaou2019}. In order to assess to what extent early weathering can keep the planet habitable despite an initial CO$_2$-rich atmosphere, in Fig. \ref{fig:initial} we test the influence of an initial atmosphere reservoirs of CO$_2$ up to 10$^{20}$ mol ($\approx$ 1 bar). We find that an initial atmosphere of 5$\times10^{19}$ mol on early Venus still allows for climate regulation and liquid water, while 1$\times10^{20}$ mol directly triggers ocean evaporation. However, our model assumed a constant planetary albedo of 0.6, and for higher initial CO$_2$ abundances the planetary albedo might increase. Altogether, if environmental conditions allowed for liquid water on an early stagnant-lid Venus shortly after magma ocean solidification, our finding of an extended early habitable period of $\approx$0.9 Gyr should remain valid, even with an initial CO$_2$-rich atmosphere in combination with a higher initial albedo.

\subsection{Plate tectonics on early Venus?}
\label{sec4b}
The idea that early Venus might have been habitable has extensively been discussed in the literature \cite{way2020,grinspoon2007,salvador2017,way2016}. Most authors consider plate tectonics when assessing a potential habitable period of early Venus. However, Venus' early tectonic regime is debated, since geological observations of Venus' surface only allow for an interpretation of the time after the global resurfacing event $\approx$500 Myr ago, while a transition from a mobile to a stagnant lid has been suggested \cite{weller2020,Gillmann:2014,head2014}. The likelihood of plate tectonics may also depend on the rheology of the mantle \cite{noack2014,oneill2007}, although a clear conclusion as to whether or not plate tectonics occurred on early Venus cannot be drawn. Our model assumes that the convective mode of Venus has been stagnant-lid throughout its entire history. It is important to emphasise that early plate tectonics would further enhance Venus' habitability, since the carbonate-silicate cycle operates more efficiently on planets where subduction of carbonates into the mantle follows a shallow temperature-depth gradient. However, our study suggests that plate tectonics is not required for extended habitability of early Venus.

\section{Conclusions}
\label{sec5}

We assessed the atmospheric evolution and habitability of early Venus accounting for weathering, carbonate burial, and decarbonation. In addition, we applied our model to Venus-sized stagnant-lid exoplanets around a Sun-like star. Our findings can be summarised as follows:

\begin{itemize}
\item[$\bullet$] As soon as weathering on stagnant-lid planets stops as a result of ocean evaporation, the supply of CO$_2$ to the atmosphere speeds up until the crust becomes depleted in carbonates. This is a consequence of a positive feedback involving atmospheric CO$_2$, surface temperature, and the rate of change of the decarbonation depth. This decarbonation feedback complements the runaway greenhouse feedback, thereby causing a dramatic rise in the surface temperature.
\item[$\bullet$] A model with an initial mantle temperature of 1900 K and an oxygen fugacity equal to the iron-w\"ustite buffer results in an atmospheric CO$_2$ similar to that observed on Venus today. If the initial atmospheric CO$_2$ reservoir allows for liquid surface water, our model accounting for weathering, burial, and decarbonation indicates a habitable time period of 900 Myr, which is 500 Myr longer than obtained from a model without weathering.
\item[$\bullet$] Stagnant-lid exoplanets should show a bimodal distribution of their atmospheric CO$_2$: Planets close to their host star should have a crust depleted in carbonates and an atmosphere abundant in CO$_2$, while planets further away from their host star could still possess an active carbon cycle restricted to their atmosphere and crust, which would efficiently regulate their surface temperature. An observation of this bimodal distribution e.g. by successfully retrieving the strong CO$_2$ fundamental band could therefore be used as a first proxy of habitability which is easier to observe than other potentially relevant factors such as additional greenhouse gases in the atmosphere.
\end{itemize}

\acknowledgments
We thank two anonymous reviewers for helpful comments and suggestions. DH was supported through the NWO StartImpuls. PB, NT and JLG acknowledge support from the DFG Priority Program SPP 1992 “Exploring the Diversity of Extrasolar Planets” (TO 704/3-1 and GO 2610/2-1). PB and NT also acknowledge support from the DFG Research Unit FOR 2440 “Matter under planetary interior conditions”. MJW was supported by NASA’s Nexus for Exoplanet System Science (NExSS). Resources supporting this work were provided by the NASA High-End Computing (HEC) Program through the NASA Center for Climate Simulation (NCCS) at Goddard Space Flight Center. MJW acknowledges support from the GSFC Sellers Exoplanet Environments Collaboration (SEEC), which is funded by the NASA Planetary Science Division’s Internal Scientist Funding Model. All relevant equations for the interior evolution model are explained in \citeA{Tosi:2017} and all model additions that adequately support reproducibility of our results are given in this paper. The GCM model is explained in \citeA{way2017}. Produced data are available in \citeA{honing:2021}.

\begin{appendix}
\section{Weathering and decarbonation on stagnant-lid planets}
\label{a1}
The global rate of seafloor weathering on Earth is usually described as the product of carbonatization per crustal volume and the global crustal production rate \cite<e.g.,>{Sleep:2001}. Applying this functional dependence to stagnant-lid planets, we assume that the silicate weathering rate of these planets $F_w$ (in mol yr$^{-1}$) depends on the atmospheric CO$_2$ partial pressure $P_{CO_2}$ and the production rate of basaltic crust $\frac{\mathrm{d}M_{cr}}{\mathrm{d}t}$ in the same manner as seafloor weathering on Earth:
\begin{linenomath}
\begin{equation}
   \frac{F_w}{F_{sfw,E}}=\frac{\left(\frac{\mathrm{d}M_{cr}}{\mathrm{d}t}\right)}{\left(\frac{\mathrm{d}M_{cr}}{\mathrm{d}t}\right)_E}\left(\frac{P_{CO_2}}{P_{CO_2,E}}\right)^\alpha,
    \label{eq:a1}
\end{equation}
\end{linenomath}
where the index $E$ denotes present-day Earth values. We can relate the present-day Earth seafloor weathering rate $F_{sfw,E}$ to the carbon ingassing rate $F_{in,E}$ into Earth's mantle:
\begin{linenomath}
\begin{equation}
   F_{in,E}=F_{sfw,E}\frac{f_E}{\xi_E},
    \label{eq:a2}
\end{equation}
\end{linenomath}
where $\xi_E$ is the fraction of seafloor weathering rate relative to the total weathering rate and $f_E$ is the fraction of buried carbonates that enters the mantle. We now assume that on present-day Earth carbon ingassing equals outgassing, which in turn is given by the product of the CO$_2$-concentration of mid-ocean ridge melt $X_E$ and the crustal production rate:
\begin{linenomath}
\begin{equation}
   F_{in,E}=X_E \left(\frac{\mathrm{d}M_{cr}}{\mathrm{d}t}\right)_E.
    \label{eq:a3}
\end{equation}
\end{linenomath}
Combining Eqs. \ref{eq:a1}--\ref{eq:a3}, we arrive at Eq. \ref{eq:weathersl} \cite<for details, see>{honing:2019}.

As in \citeA{honing:2019}, we track carbonated crust that is buried by new lava flows and thereby sinks downward. Neglecting heat sources in the crust, we calculate the temperature $T(z)$ at depth $z$ assuming a linear temperature profile:
\begin{linenomath}
\begin{equation}
     T(z)=T_s+\frac{z(T_m-T_s) }{D_l+\delta_m },
    \label{eq:a4}
\end{equation}
\end{linenomath}
where $T_s$ and $T_m$ are the surface and mantle temperatures, and $\delta_m$ and $D_l$ are the thicknesses of the thermal boundary layer and of the stagnant lid, respectively. The decarbonation temperature is assumed to increase linearly with depth:
\begin{linenomath}
\begin{equation}
     T_{decarb}=Az+B,
    \label{eq:a5}
\end{equation}
\end{linenomath}
where $A$ and $B$ are constants from \citeA{Foley:2018}. Combining Eqs. \ref{eq:a4} and \ref{eq:a5}, we arrive at Eq. \ref{eq:zdecarbsl}.

\end{appendix}

\bibliography{hoening21.bib}

\end{document}